\documentclass[prl,twocolumn,superscriptaddress,showpacs,nobalancelastpage]{revtex4}
\usepackage{graphicx}

\begin{document}

\title{Magneto-transport through graphene nano-ribbons}

\author{Jeroen B. Oostinga}
\affiliation{DPMC and GAP, University of Geneva, 24 quai Ernest
Ansermet, CH-1211 Geneva, Switzerland}
\affiliation{Kavli Institute of NanoScience, Delft University of
Technology, PO Box 5046, 2600 GA Delft, The Netherlands}

\author{Benjamin Sac\'{e}p\'{e}}
\affiliation{DPMC and GAP, University of Geneva, 24 quai Ernest
Ansermet, CH-1211 Geneva, Switzerland}

\author{Monica F. Craciun}
\affiliation{Centre for Graphene Science, Department of Engineering,
Mathematics and Physical Sciences, University of Exeter, North Park
Road, Exeter, EX4 4QF, United Kingdom}

\author{Alberto F. Morpurgo}
\affiliation{DPMC and GAP, University of Geneva, 24 quai Ernest
Ansermet, CH-1211 Geneva, Switzerland}

\date{\today}

\begin{abstract}

We investigate magneto-transport through graphene nano-ribbons as a
function of gate and bias voltage, and temperature. We find that a
magnetic field systematically leads to an increase of the
conductance on a scale of a few tesla. This phenomenon is
accompanied by a decrease in the energy scales associated to
charging effects, and to hopping processes probed by
temperature-dependent measurements. All the observations can be
interpreted consistently in terms of strong-localization effects
caused by the large disorder present, and exclude that the
insulating state observed in nano-ribbons can be explained solely in
terms of a true gap between valence and conduction band.

\end{abstract}


\maketitle

Single-layer graphene is a zero-gap semiconductor, whose valence and
conduction bands touch at two inequivalent points (the K and K'
points) at the edge of the first Brillouin zone \cite{Gei}. Owing to
the absence of a band-gap, the conductance of graphene remains
finite irrespective of the position of the chemical potential. The
impossibility to turn off the conductivity hinders the fabrication
of high-quality field-effect transistors, and poses problems for the
use of graphene in electronic devices. A possible solution is the use of graphene nano-ribbons,
in which the opening of a gap between valence and conduction band has been predicted theoretically \cite{Gap-th}.

According to theory, a gap in nano-ribbons should open as a consequence of size quantization and,
depending on the specific case considered, interaction effects
\cite{Gap-th}. Transport experiments indeed find that the
conductance through narrow ribbons is very strongly suppressed when
the Fermi energy is close to the charge neutrality point
\cite{Kim-Gap,Gap-exp}. However, it is unclear whether the
opening of a band-gap is the correct explanation for the
experimental observations. On the one hand, a gap would lead to a
conductance suppression similar to what is found experimentally,
with characteristic energy scales (as probed in temperature- and
bias-dependent measurements) comparable to the
observed ones \cite{Kim-Gap}. On the other hand, the theoretical
predictions rely on edge structures that are ideal on the atomic
scale, whereas this is certainly not the case in real devices. This
is why many theoretical studies have analyzed the role of strong
localization of electron wave-functions --possibly in combination
with charging effects-- as the origin of the conductance suppression
observed in the experiments on nano-ribbons \cite{SL-th}.

Existing studies of transport through graphene nano-ribbons
have mainly relied on measurements of conductance as a function of carrier density and temperature,
and have not led to a definite conclusion as to the nature of the transport gap.
Here, we present a systematic investigation focusing on the magneto-transport properties. We show that, in the transport gap, the application of magnetic field always leads to a conductance increase, together with
a decrease in the characteristic energy scale associated to the conductance suppression.
We perform an analysis of these measurements that excludes the opening of a band-gap as the sole explanation for the observed insulating behavior, and accounts for all our observations (including
the gate-voltage and temperature dependence) in a consistent way,
in terms of strong-localization effects caused by the larger disorder present in narrow ribbon geometries.

\begin{figure}[!t]
\includegraphics[width=3.4in]{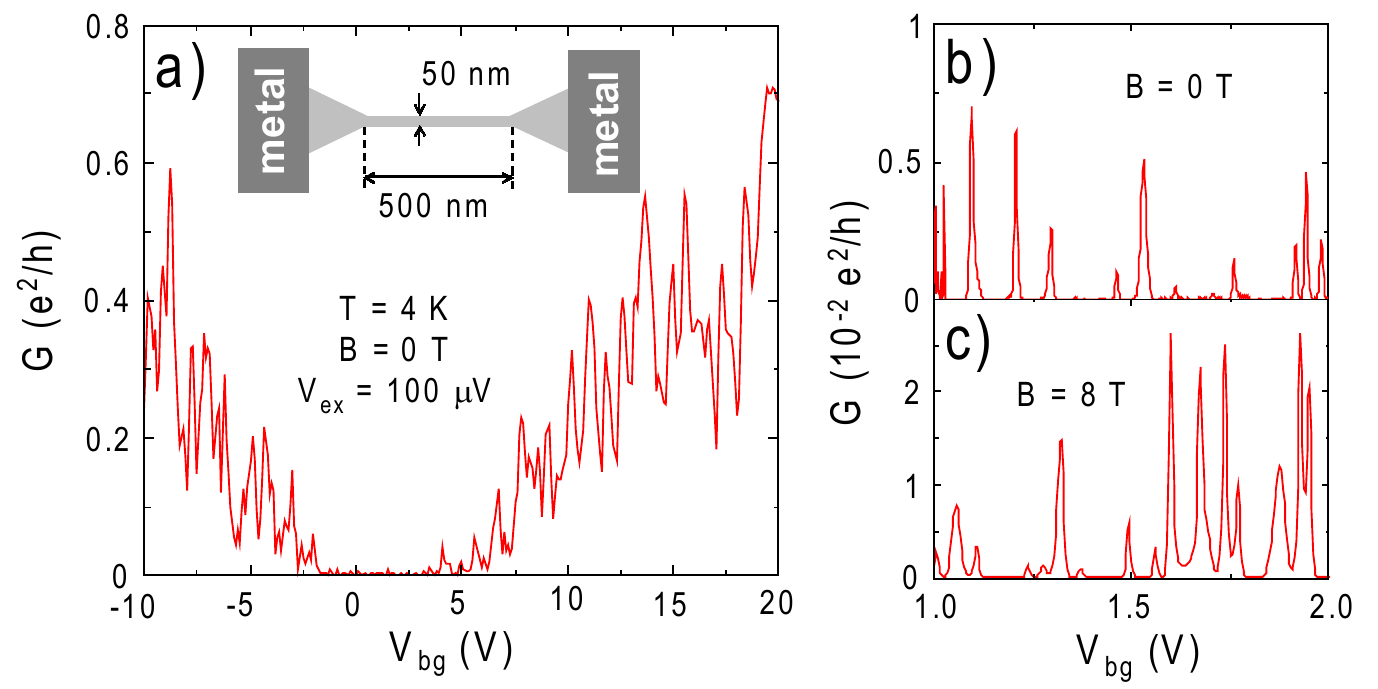}
\caption{(a) Conductance $G$ as a function of the gate-voltage
$V_{bg}$, showing a transport gap at low charge densities. The inset
shows a scheme of the device geometry (the substrate is highly doped silicon covered by a 285 nm
SiO$_2$ layer). Panel (b) and (c) zoom-in on the
Coulomb-blockade peaks present in the transport gap at $B=0$ and 8
T, respectively. } \label{Fig1}
\end{figure}

We have investigated many devices consisting of graphene
nano-ribbons fabricated using the by-now conventional exfoliation
procedure. The details of the device fabrication process and ribbon
patterning (based on etching in an Ar plasma) are identical to those
that we have used for the realization of Aharonov-Bohm rings (see
\cite{AB}). The device geometry is schematically shown in
the inset of Fig. 1(a). Different dimensions were used, with the
ribbon width $W$ varying approximately between 50 and 100 nm, and
the ribbon length $L$ between 500 and 1000 nm. The region where the
metal electrodes contact the graphene layer is intentionally kept
large, to exclude effects due to the contact resistance. Conductance
measurements on many different devices yielding consistent results
were performed in a two-terminal configuration as a function of
back-gate voltage, temperature, and magnetic field. Unless stated
otherwise, here we show data obtained from one of these ribbons
(with length $L=500$ nm and width $W=50$ nm), representative of the
overall behaviour observed.

Fig. 1(a) shows the linear conductance $G$ of the ribbon as a
function of gate voltage. The conductance is strongly suppressed in
the voltage range $[-2 V, +4 V]$, that we refer to as the transport
gap (the transport gap depends weakly on the ribbon length and it is
larger for longer devices, consistently with the results of Ref. \cite{Kim-SL}).
Fig. 1(b) zooms-in on this range, showing conductance peaks
at several gate voltages (with the conductance vanishing elsewhere)
originating from charging effects. Measurements as a function of
gate ($V_{bg}$) and bias ($V_{sd}$) voltage show that the
conductance is highly suppressed at low $V_{sd}$ (Fig. 2(a)), and
exhibit characteristic "diamond-like" structures (Fig. 3(a)) typical
of Coulomb blockade. These results are similar to others already
reported \cite{Kim-Gap,Gap-exp}, and clearly indicate that the
electrons are confined in small regions of the ribbon --which we
call islands-- whose area is sufficiently small to have a
significant charging energy. Fig. 3(a) shows that the diamonds are
irregular and partially overlap, indicating that the islands in the
ribbon vary in size, and that electrons have to propagate through
several of them while traversing the ribbon. At a microscopic level,
the question is what is the mechanism that determines the presence
of the islands, i.e. the presence of a band-gap or the effect of
strong localization.

\begin{figure}[!t]
\includegraphics[width=3.4in]{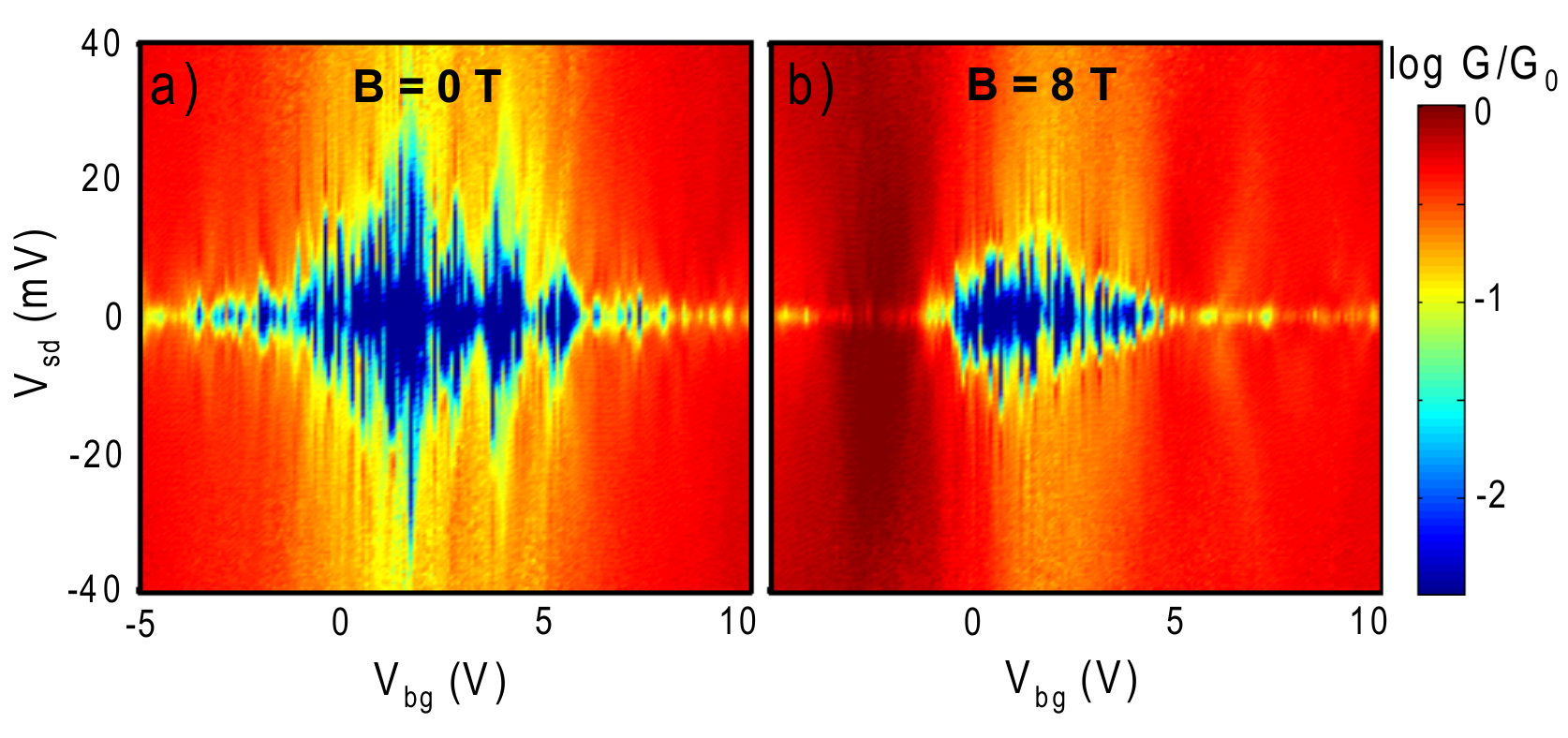}
\caption{Panels (a) and (b) show the color plots of $\log(G)$ as a
function of gate ($V_{bg}$) and bias ($V_{sd}$), respectively at
$B=$ 0 and 8 T (the measurements are taken at $T=$ 4.2 K;
$G_0=e^2/h$). The comparison of the two panels shows how in a
magnetic field the extent of the high-resistance part is suppressed.
} \label{Fig2}
\end{figure}

The analysis of the influence of magnetic field provide additional
important information. In Fig. 2 (a) and (b), the
conductance measured as a function of gate and bias voltage at $B=$
0 T and 8 T can be directly compared. The bias
and gate voltage ranges where the conductance is suppressed are
significantly smaller when a high magnetic field is applied. Also
the Coulomb peaks and diamonds are affected
(compare Fig. 1(b) and (c), and Fig. 3(a) and (b)): in the presence
of a high magnetic field, the average number of Coulomb peaks in a
given gate voltage range is larger, and the average peak conductance
is higher. Qualitatively, these observations indicate that the
average island size increases when a magnetic field field is
applied, since for larger islands the charging energy is smaller.

The continuous evolution from $B=0$ T to 8 T is shown in Fig. 4(a).
In Fig. 4(b) we show the average (over gate voltage varying in the
transport gap) of the conductance as a function of magnetic field.
Since the conductance between the Coulomb peaks is negligibly small,
this quantity measures the average peak conductance and the number
of peaks present. Fig. 4(b) shows the average conductance for the
sample that we are discussing, and for two other nano-ribbons having
width $W=80$ nm and $W=110$ nm. We see that the
average conductance increases with magnetic field for $B<3$ T and
that it saturates for $B>3$ T. We conclude that the average number
of Coulomb peaks and the peak conductance increase for $B<3$ T, but
remains approximately constant for $B>3$ T. Unexpectedly, the
average conductance saturates at approximately similar values of the
magnetic field in the different devices, irrespective of the ribbon
width.

In order to analyze the results of the magneto-transport
experiments, we first go back to the data shown in Fig. 1(a), and
look at a larger gate voltage range (outside the transport gap),  to
estimate the value of the electron mean-free path. From the average
slope (i.e., neglecting the fluctuations) of the conductance versus
gate voltage, $\partial G/\partial V_{bg}$, measured at high charge
density (where the conductance is much larger than in the transport
gap), we determine the field-effect mobility
$\mu_{FET}=\frac{tL}{W\epsilon_r\epsilon_0}\frac{\partial
G}{\partial V_{bg}}\approx 1000$ cm$^2$/Vs. This value is smaller
than the typical values usually found for large graphene flakes on
SiO$_2$ substrates ($\mu \sim 5000-10000$ cm$^2$/Vs \cite{Gei}),
confirming that graphene nanoribbons are more disordered
\cite{note1}. We determine $l_m$ using the Einstein relation
$\sigma=\nu e^2D$, where $D$ is the diffusion constant and $\nu$ the
density of states
($\nu(\varepsilon_F)=8\pi|\varepsilon_F|/(h^2v_F^2)$, with
$\varepsilon_F$ and $v_F$ the Fermi energy and velocity,
respectively). From the conductivity, $\sigma\approx
G_{square}=GL/W$ measured at $V_{bg}=-10$ V, and determining
$\varepsilon_F$ from the charge density induced electrostatically
$n(V_{bg}=-10V)=4\pi\varepsilon_F^2/(h^2v_F^2)\approx 1\cdot
10^{12}$ cm$^{-2}$, we obtain $D\approx 0.006$ m$^2$/s. The
corresponding electron mean-free path is then obtained from
$l_m=2D/v_F\approx 10$ nm,  smaller than the ribbon width ($W=50$
nm).

\begin{figure}[!t]
\includegraphics[width=3.4in]{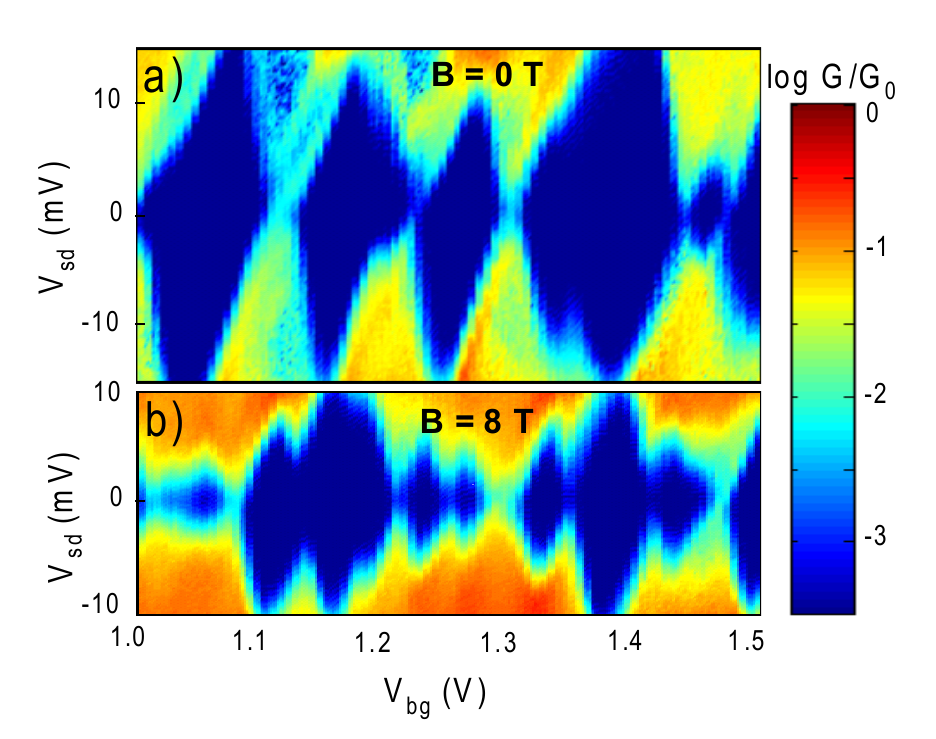}
\caption{Coulomb diamonds measured in the transport gap at $T=$4.2
K, at $B=0$ T (a) and $B=8$ T (b). The typical height $\Delta
V_{sd}$ and width $\Delta V_{bg}$ of the diamonds at high field are
smaller than at zero field.} \label{Fig3}
\end{figure}

The short mean-free path values are indicative of strong disorder,
and suggest that a scenario based on strong-localization may be
appropriate (consistently with the conclusion reported recently in
Ref. \cite{Kim-SL}, and theoretical predictions \cite{SL-th}). A
signature of strong localization is a positive magneto-conductance,
originating from an increase of the localization length due to time
reversal symmetry breaking \cite{Books,Ger}. In the strong
localization regime, time reversal symmetry is broken when the
magnetic flux through an area corresponding to the square of the
localization length $\xi^2$ is of the order of the flux quantum
$h/e$. This mechanism can explain our data of Fig. 4(b), which show
that the average conductance at low $n$ does indeed increase when
the magnetic field is increased from 0 T to $\sim$3 T, and provide
an estimate of the localization length of $\xi \approx (h/eB)^{1/2}
\approx 35-40$ nm. This value of $\xi$ is somewhat smaller than (but
comparable with) the width of the three ribbons whose data are
plotted in Fig. 4(b): this is why the magneto-conductance saturates
at approximately the same magnetic field in the different devices.
Further, we estimate the phase coherence length
$l_\varphi=(D\tau_{\varphi})^{1/2}$ by using the value of the
diffusion constant obtained above and the phase coherence time
$\tau_\varphi\approx5$ ps reported in Ref. \cite{Sav}. We obtain
$l_\varphi\approx 175$ nm, from which we conclude that
$l_m<\xi<l_\varphi$, the correct hierarchy of length scales for the
strong localization regime \cite{Books}. Note also that at low
charge density, where the number of transverse channels occupied is
small, and $\xi \sim l_m$, as expected from theory \cite{Books}. The
increase in conductance observed at higher charge density can then
be attributed to an increase in localization length, due to the
increase in the number of occupied transverse channels.

\begin{figure}[!t]
\includegraphics[width=3.4in]{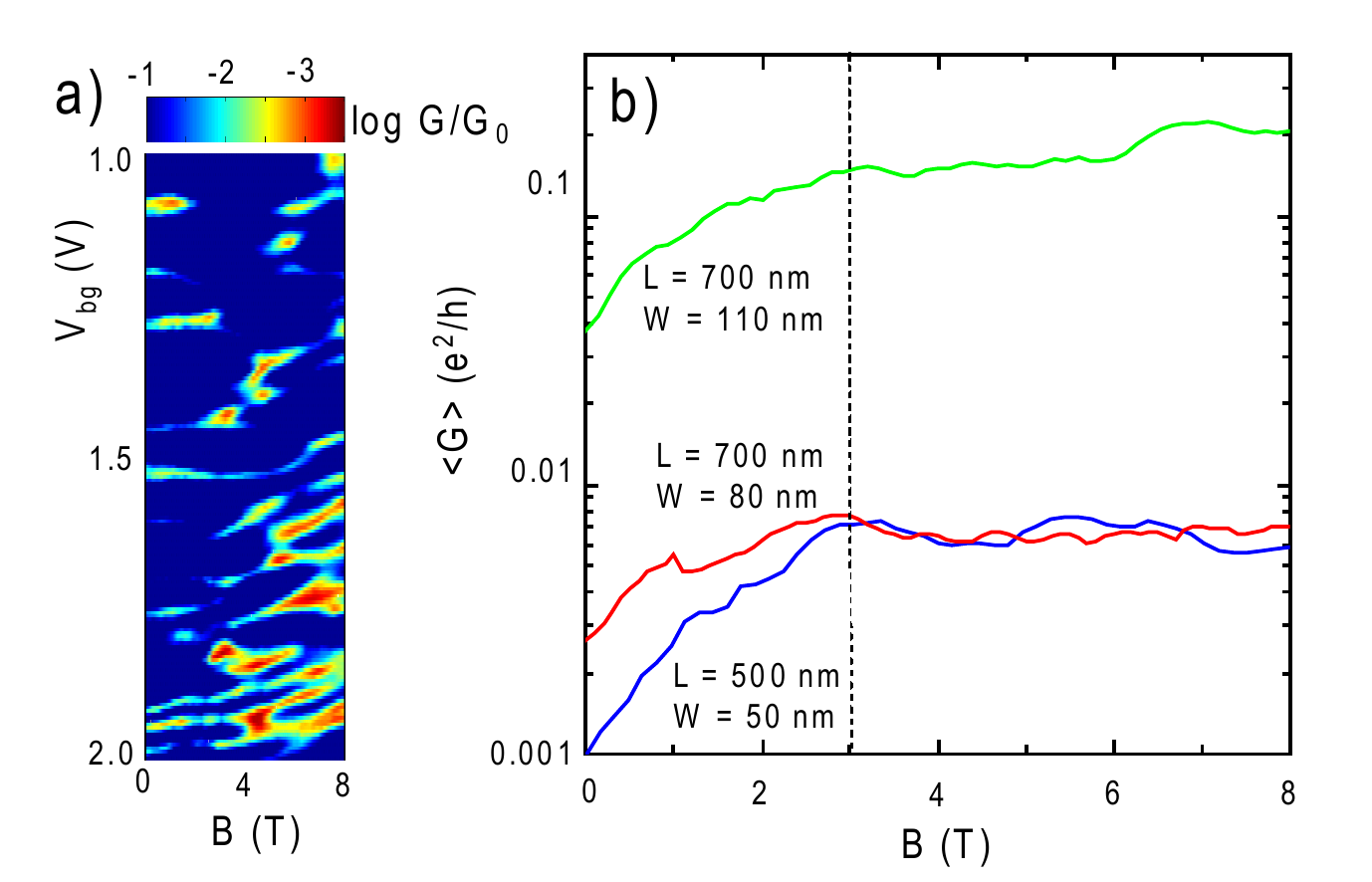}
\caption{(a) Color plot of $\log(G)$ as a function of $V_{bg}$ (in
the transport gap) and of $B$ (measurements done at $T=$ 4.2 K). The
average peak conductance and the number of Coulomb peaks increase
with increasing magnetic field (consistent with Fig. 1(b,c) and Fig.
3). Panel (b) shows the average of conductance over gate voltage, as
a function of magnetic field. Noticeably, this quantity increases
with increasing $B$ up to approximately 3 T (irrespective of the
ribbon width), and saturates for higher field values. } \label{Fig4}
\end{figure}

It is particularly important to discuss whether the observed
positive magneto-conductance can be understood if the formation of
islands in the ribbons is due to the opening of a band-gap (and not
by strong-localization effects). In the presence of a high magnetic
field, well-defined Landau levels are expected to appear in the
energy spectrum when the magnetic length, $l_B=(\hbar/eB)^{1/2}$
becomes sufficiently smaller than the ribbon width (when $l_B << W$,
the finite width does not affect the Landau levels). Owing to the
presence of a zero-energy Landau level in graphene \cite{QHE}, a
large magnetic field should cause the closing of the gap and lead to
the appearance of edge states that would contribute to the
conductance ($\sim e^2/h$). This mechanism could explain a positive
magneto-conductance. In the experiments, however, we do not observe
any signature of the formation of edge states --the low-temperature
conductance in the transport gap remains always much smaller than
$e^2/h$-- despite the fact that $l_B\approx$ 10 nm for $B=$ 8 T (ten
time smaller than the largest ribbon that we discuss here). This
observation again implies that ribbons are highly disordered: even
for the widest ribbon disorder is strong enough to cause
backscattering between counter-propagating states at opposite edges,
leading to their localization. In other words, even if we attempt to
explain the positive magneto-conductance in terms of a band gap, we
are forced to conclude that localization effects are nevertheless
dominating \cite{note2}.

We now look at the temperature dependence of the conductance
measured in the transport gap. For strongly localized electrons,
transport occurs via either nearest neighbor hopping (NNH) or
variable range hopping (VRH) \cite{Books}, and the latter mechanism
dominates at low temperature. Fig. 5 shows the measurement of the
gate-averaged conductance in the transport gap as a function of
temperature in the range $2-60$ K, for  $B=0$ T and $B=8$ T. In case
of NNH, the temperature dependence of the conductance should be
proportional to $G \propto \exp -(T_0/T)$. However, the plot shown
in the inset of Fig. 5 indicates that Arrhenius law does not
reproduce the data well. Rather, the data fits better to $G \propto
\exp -(T_0/T)^{1/2}$, which is the temperature dependence expected
for VRH (see Fig. 5). The exponent $n=1/2$ corresponds to
two-dimensional hopping in the presence of Coulomb interaction
(which is expected given the obvious presence of Coulomb-blockade
effects), or for quasi one-dimensional hopping with or without
interactions. This behavior indicates that electronic transport is
dominated by variable-range hopping.

The parameter $T_0$ is related to the energy needed by an electron
to hop between localized states. From the linear fit to the data of
Fig. 5  we obtain $kT_0 \approx 6$ meV at $B=0$ T and a smaller
value, $kT_0 \approx 4$ meV at $B=8$ T when time reversal symmetry
is broken. This result agrees with what we would expect if $\xi$
is larger at high magnetic field. Indeed, the
relevant energy scales of the localized states involved in the
hopping are the single-particle level spacing and the charging
energy, which are inversely proportional to the extension of the
wave-function of the localized states (i.e., they decrease when
$\xi$ increases). The values found for $T_0$ are, as expected for
the variable range hopping regime, a fraction of the characteristic
values of single level spacing and charging energy, which, depending
on the detail of the estimate range from about 10 meV to a few tens
of meV (at $B=0$ T, with $\xi \simeq 35-40$ nm).

\begin{figure}[!t]
\includegraphics[width=2.8in]{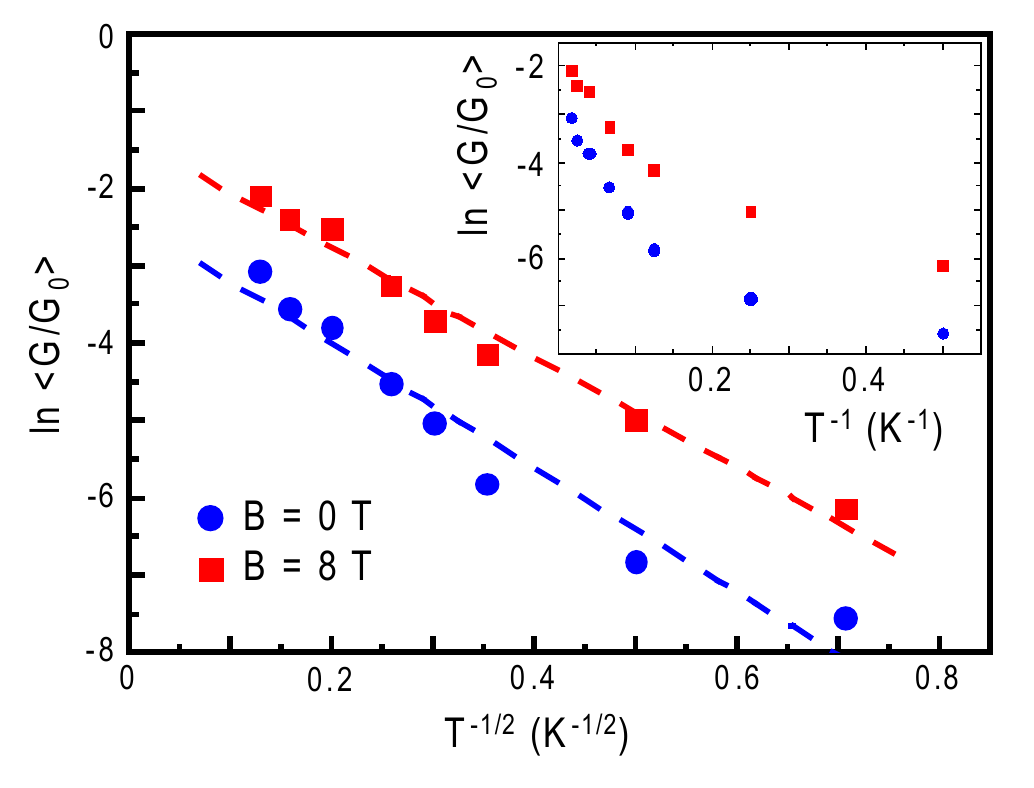}
\caption{The average conductance $<G>$ in the transport gap as a
function of $T^{-1/2}$ at $B=$ 0 T and $B=$ 8 T. The dashed lines
are linear fits to the data. In the inset, the same data is plotted
as a function of $T^{-1}$.}
\label{Fig5}
\end{figure}

In summary, magneto-transport experiments provide a consistent picture
as to the origin of the transport gap observed in graphene
nano-ribbons: they indicate that strong-localization effects --and
not a gap between valence and conductance band-- are the cause of
the observed insulating state that is observed experimentally.

\begin{acknowledgments}

We thank Y. Blanter, L. Vandersypen, M. Fogler for
useful discussions, and S. Russo for support in fabricating
several devices. This work was supported by the NWO VICI,
NanoNed, the Swiss National Science Foundation (project
200021-121569), and by the NCCR MaNEP.

\end{acknowledgments}


\begin{references}
\bibitem{note2}
At 8 tesla the energy gap between the lowest Landau levels is much
larger than the expected band-gap: since features associated to
Landau levels are not visible, the much smaller band-gap should also
not be expected to determine the device properties.
\end{references}
\end{document}